\documentclass[twocolumn,epjc3]{svjour3}
\smartqed  
\RequirePackage{graphicx}
\begin{document}
\title{An alternative response to the off-shell quantum fluctuations:
A step forward in resolution of the Casimir puzzle}

\titlerunning{An alternative response to the off-shell quantum fluctuations
}

\author{
G.~L.~Klimchitskaya\thanksref{addr1,addr2}\and
V.~M.~Mostepanenko\thanksref{addr1,addr2,addr3,e1}}

\authorrunning{G.~L.~Klimchitskaya and V.~M.~Mostepanenko}

\thankstext{e1}{e-mail: vmostepa@gmail.com (corresponding author)}
\institute{
Central Astronomical Observatory
at Pulkovo of the Russian Academy of Sciences,
St.Petersburg, 196140, Russia \label{addr1}
\and
 Institute of Physics, Nanotechnology and
Telecommunications, Peter the Great Saint Petersburg
Polytechnic University, Saint Petersburg, 195251, Russia
\label{addr2}
\and
Kazan Federal University, Kazan, 420008, Russia
\label{addr3}
}

\date{Received: 13 August 2020 / Accepted: 10 September  2020}
%
\maketitle

\abstract{
The spatially nonlocal response functions are proposed which nearly
coincide with the commonly used local response for electromagnetic
fields and fluctuations on the mass shell, but differ significantly
for the off-shell fluctuating field. It is shown that the fundamental
Lifshitz theory using the suggested response functions comes to an
agreement with the measurement data for the Casimir force without
neglecting the dissipation of free electrons. We demonstrate that
ref\-lec\-tan\-ces of the on-shell electromagnetic waves calculated using
the nonlocal and commonly employed local responses differ only
slightly. The Kramers-Kronig relations for nonlocal response
functions possessing the first- and second-order poles at zero
frequency are derived, i.e., the proposed response satisfies the
principle of causality. An application of these results to resolution
of the Casimir puzzle, which lies in the fact that the Lifshitz theory
is experimentally consistent only with discarded dissipation, is
discussed.
} 
\newcommand{\ri}{{\rm i}}
\newcommand{\il}{{{\rm i}\xi_l}}
\newcommand{\kb}{{k_{\bot}}}
\newcommand{\kk}{{({\rm i}\xi_l,k_{\bot})}}
\newcommand{\kTl}{{k_l^T}}
\newcommand{\vk}{{\boldmath$k$}}
\newcommand{\eps}{{\varepsilon}}
\newcommand{\teps}{{\tilde{\varepsilon}}}
\newcommand{\rTM}{{r_{\rm TM}}}
\newcommand{\rTE}{{r_{\rm TE}}}
\newcommand{\zTM}{{Z_{\rm TM}}}
\newcommand{\zTE}{{Z_{\rm TE}}}
\section{Introduction}

Beginning in 2000, much attention is being given to the Casimir
force \cite{1} acting between closely spaced uncharged surfaces. This
force is caused by the quantum fluctuations (both zero-point
and thermal) of the electromagnetic field. It extends
familiar van der Waals force \cite{2} to larger separations where
the relativistic effects become essential. The
general theory of van der Waals and Casimir forces between two material
plates developed by Lifshitz \cite{3} is in fact semiclassical.
It describes electromagnetic fluctuations in the framework of thermal
quantum field theory in the Matsubara formulation, but the response of
matter to these fluctuations is treated classically by means of the
standard continuity boundary conditions where
the frequency-dependent dielectric permittivity plays the role of a
response function. Taking into consideration that the
Casimir effect finds numerous multidisciplinary applications in
quantum field theory, physics of elementary particles, gravitation and
cosmology, atomic physics, condensed matter physics, as well as in
nanotechnology (see, e.g., the monographs \cite{4,5,6,7,8,9,10}), it is
hardly surprising that the Lifshitz theory was used and cited in
thousands of papers.

Over a protracted period of the last 20 years, the Lifshitz theory
has been facing a challenge when calculating the Casimir force
between metallic surfaces and when comparing the results obtained with the
measurement data. According to \cite{11}, at room temperature
this theory predicts anomalously large thermal effect in the Casimir force
even at relatively short separations below 1 $\mu$m if the electromagnetic
response of a metal is described by means of the dissipative
Drude dielectric permittivity
(or is obtained from the available optical data extrapolated by means
of the Drude permittivity down to zero frequency). The measurement data of
numerous precise experiments excluded this prediction and were found
in good agreement with the Lifshitz theory if the electromagnetic
response of a metal is described using the optical data of
a metal extrapolated by the dissipationless plasma permittivity
 (see a review of the first experiments by R.\ S.\ Decca in \cite{24})
and later experiments
\cite{16,17,18,19,20,21,22,23}).
The situation in the field was also reviewed in \cite{25}.

Many attempts have been made to explain the about 5\%
disagreement between the Lifshitz theory and the measurement data
of \cite{24,16,17,18,19}
with the role of some unaccounted effects in the surface roughness
\cite{26,27}, variation of the optical properties of Au films
\cite{28}, patch potentials \cite{29,30}, and by deviations from the
proximity force approximation used in computations when one of the
plates is replaced with a sphere \cite{31,32,33,34,35}.
The facts have been conclusively established by the
differential force measurement of   \cite{20}, where the
predictions of the Lifshitz theory using the Drude and the plasma
responses differ by up to a factor of 1000. In this experiment, the
Lifshitz theory using the Drude response function
for calculation of the Casimir force was ultimately excluded over
the separation range from 200 to 700 nm, whereas the same theory
using the plasma response was found to be in good
agreement with the measurement data. Later it was shown that the
same is true up to the $1.1~\mu$m separation \cite{21,22,23}.

This situation presents a puzzle \cite{36} when it is considered that
the Lifshitz theory is based on the first principles of quantum
electrodynamics at nonzero temperature and the Drude response
function takes proper account of the dissipation of conduction
electrons, whe\-re\-as the plasma response excludes this
phenomenon from consideration and is, in fact, applicable only
at high frequencies. In the only experiment found in better agreement
with the Drude response function
at separations of a few micrometers \cite{37}, not the Casimir force
itself was measured, but up to an order of magnitude larger force of
unknown origin. As shown in  ~\cite{38}, the interpretation of
this experiment is, in fact, uncertain.

It is no less surprising that for metals with perfect crystal
lattices the Lifshitz theory violates the third law of thermodynamics
(the Nernst heat theorem) when the Drude response is used but is
thermodynamically consistent when using the plasma response
\cite{41,42,43,44,45}. It has been shown that for lattices with
impurities the
Casimir entropy calculated using the Drude response satisfies the Nernst
theorem \cite{46,47,48}. However, for a perfect crystal lattice, which
is the basic theoretical model of condensed matter physics, the thermodynamic
problem remains unresolved.

In the absence of a plausible resolution for this puzzle over a
long period, it was hypothesized \cite{23} that the problem might
be caused by an incomplete understanding of the response of metal
to quantum fluctuations. The point is that the Lifshitz formula for
the Casimir pressure (see Sect.~2) depends on the response of a plate
material to the quantum fluctuations both on the mass shell and
off the mass shell. The spatially local
Drude response function is not as fundamental as
the Lifshitz theory. It is derived in the framework of
kinetic theory and  Kubo formula  on several assumptions, such as the
zero wave vector limit and the relaxation time approximation, which
means that the current-current correlation function exponentially
decays in time. The Drude function describes reasonably well the response
of metal to real electromagnetic field on the mass shell.
It is worth mentioning, however,  that an experimentalist cannot irradiate
a metallic film by the off-shell electromagnetic fluctuations and measure
the real and imaginary parts of its complex index of refraction. Because of
this, there are neither exact theoretical nor direct experimental
justifications of an assertion that the response of metals to electromagnetic
fluctuations off the mass shell is described by the Drude function.

In this regard, of fundamental
interest is graphene governed by the Dirac model \cite{49,50}.
The polarization tensor of this 2D material was found exactly on
the basis of first principles of thermal quantum field theory in
the Matsubara formulation \cite{51,52,53,54}. The resulting
 response of graphene to the electromagnetic field is
spatially nonlocal, i.e.,
 is described by the two
functions depending on both the frequency and the wave vector.
The Casimir force in graphene systems, calculated using these
functions in the framework of the Lifshitz theory \cite{55}, is
in good agreement with the measurement data \cite{56}, and the
Casimir entropy satisfies the Nernst heat theorem \cite{59,60}.
It seems justified to check the possibility of something similar for metals,
although the exact polarization tensor for 3D metallic bodies is,
of course, unattainable.

In this paper, we propose the spatially nonlocal
phenomenological response functions, which demonstrate nearly the same
response as the standard Drude function to the electromagnetic field on the mass
shell, but lead to quantitative differences for the  off-shell
fluctuating field. The response functions suggested here take into account the
dissipation of conduction electrons and simultaneously
lead to as good agreement with the available measurement
data for the Casimir force  as does the
plasma response. At separations of a few micrometers, the force values,
calculated here by using the suggested nonlocal response, are sandwiched between
the theoretical predictions obtained using the standard Drude
and the plasma responses. We demonstrate that the suggested response satisfies
the Kramers-Kronig relations for functions possessing the first- and second-order
poles at zero frequency, i.e, it is in agreement with the principle of
causality. It also leads to experimentally indistinguishable differences
in the reflectances of electromagnetic waves on the mass shell incident
on a metallic surface, as compared to the standard Drude response.

The paper is organized as follows: In Sect.~2, we present the summaries of the
standard, local, Lifshitz theory and the Lifshitz theory employing the
spatially nonlocal response functions. In Sect.~3, the phenomenological
nonlocal  response functions are introduced which produce
an alternative response to the electromagnetic fluctuations off the mass
shell. In Sect.~4, we demonstrate that the proposed nonlocal response functions
bring the Lifshitz theory in agreement with the measurement data for the
Casimir force. Section~5 demonstrates that the suggested nonlocal
 response leads to nearly the same reflectances of the electromagnetic wa\-ves
on the mass  shell as does the standard
Drude function. In Sect.~6, we prove that the proposed nonlocal response is causal
and satisfies the Kramers-Kronig relations. In Sect.~7, the reader will find
our conclusions and a discussion.

\section{The Lifshitz theory with spatially local and nonlocal response functions}

According to the Lifshitz theory, the
Casimir pressure between two thick metallic plates (semispaces) in
thermal equilibrium with the environment at temperature $T$ is given by
the Lifshitz formula \cite{3}
\begin{equation}
P(a)=-\frac{k_BT}{\pi}\!\sum_{l=0}^{\infty}{\vphantom{\sum}}^{\!\prime}\!\!\!
\int_0^{\infty}\!\!\!\!\!q_l\kb d\kb \!\sum_{\alpha}\!\left[
\frac{e^{2aq_l}}{r_{\alpha}^{2}\kk }-1\right]^{-1}\!\!\!\!\!\!\!,
\label{eq1}
\end{equation}
\noindent
where $k_B$ in the Boltzmann constant, $\kb$ is the magnitude of the
projection of wave vector {\vk} on the plane of plates,
$q_l=(k_{\bot}^2+\xi_l^2/c^2)^{1/2}$,
$\xi_l=2\pi k_BTl/\hbar$, $l=0,\,1,\,2,\,\ldots$ are the Matsubara frequencies,
and the summation in $\alpha$ is over the transverse magnetic (TM) and
transverse electric (TE) polarizations of the electromagnetic field
(the prime on the summation sign divides the term with $l=0$ by 2).
The frequently used notation $\kb$ is due to the fact that the wave vector
projection on the plane of plates is perpendicular to the Casimir force which
is aligned with the $x_3$ axis.

In the standard Lifshitz theory, the reflection coefficients $r_{\alpha}$ have
the Fresnel form and are expressed via the frequency-dependent dielectric
permittivity $\eps_l=\eps(\il)$ of the plate material at the imaginary
Matsubara frequencies
\begin{equation}
\rTM(\il,\kb)=\frac{\eps_lq_l-k_l}{\eps_lq_l+k_l},
\quad
\rTE(\il,\kb)=\frac{q_l-k_l}{q_l+k_l},
\label{eq2}
\end{equation}
\noindent
where $k_l=(k_{\bot}^2+\eps_l\xi_l^2/c^2)^{1/2}$.
The permittivity $\eps_l$ describes the response of a metal to the electromagnetic
field. It can be found from the measured imaginary part of the complex index of
refraction using the Kramers-Kronig relations \cite{10}.

Taking into account that the optical data for the complex index of
refraction are available only to some minimum energy (e.g., to 0.1~eV for Au
\cite{61}), at lower energies the data are usually extrapolated by means
of the Drude function
\begin{equation}
\eps_D(\omega)=1-\frac{\omega_p^2}{\omega(\omega+\ri\gamma)},\quad
\eps_D(\il)=1+\frac{\omega_p^2}{\xi_l(\xi_l+\gamma)},
\label{eq3}
\end{equation}
\noindent
where $\omega_p$ is the plasma frequency and $\gamma$ is the relaxation parameter.
As discussed in Sect.~1, this approach leads to contradictions with measurements
of the Casimir force. The plasma response function is obtained from (\ref{eq3})
by putting $\gamma=0$
\begin{equation}
\eps_p(\omega)=1-\frac{\omega_p^2}{\omega^2},\quad
\eps_p(\il)=1+\frac{\omega_p^2}{\xi_l^2}.
\label{eq4}
\end{equation}
\noindent
When it is used for an extrapolation of the optical data, the Lifshitz theory is
brought in agreement with experiments on measuring the Casimir force.

The sharp distinction between the response functions (\ref{eq3}) and (\ref{eq4}) is
in the values of the TE reflection coefficient at zero Matsubara frequency.
In the case of the Drude response (\ref{eq3}), we have from (\ref{eq2})
\begin{equation}
r_{{\rm TM},D}(0,\kb)=1,\quad r_{{\rm TE},D}(0,\kb)=0,
\label{eq5}
\end{equation}
\noindent
whereas the plasma response (\ref{eq4}) leads to
\begin{eqnarray}
&&
r_{{\rm TM},p}(0,\kb)=1,
\nonumber \\
&&
r_{{\rm TE},p}(0,\kb)=
\frac{\kb c-\sqrt{k_{\bot}^2c^2+\omega_p^2}}{\kb c+
\sqrt{k_{\bot}^2c^2+\omega_p^2}}.
\label{eq6}
\end{eqnarray}

Just this distinction results in a disagreement between experiment and theory
when the Drude response is used for extrapolation of the optical data
and in agreement when an
extrapolation is made by means of the plasma response.
The crucial point in the above is that the Casimir pressure (\ref{eq1}) is
determined by the electromagnetic fluctuations on the mass shell, for which
$\kb\leq\xi_l/c$, and also off the mass shell for which $\kb>\xi_l/c$.
In so doing a common response function $\eps_l$ is used for both
types of fluctuations which are often called the propagating and evanescent
waves, respectively. We emphasize that the electromagnetic field in (\ref{eq5})
and (\ref{eq6}) is characterized by $\xi=0$, $\kb>0$ and, thus, is just
off the mass shell.

During the last years, the Lifshitz theory was generalized for the bodies made
of any material and of arbitrary geometrical shape \cite{62,63,64,65}.
It was shown that in the case of plane-parallel configurations the Casimir
pressure preserves its form (\ref{eq1}), but the reflection coefficients
may be quite different from the Fresnel ones (\ref{eq2}).
Specifically, in the presence of spatial dispersion
the response of metal to the electromagnetic field is described by
a tensor which is determined by the two independent functions --- the
longitudinal, $\eps^L(\omega,\mbox{\vk})$, and transverse,
$\eps^T(\omega,\mbox{\vk})$, dielectric permittivities depending on the frequency
$\omega$ and wave vector {\vk} \cite{66,67}. Here, the longitudinal and transverse
electric fields are parallel and perpendicular to {\vk}, respectively.

With account of spatial dispersion the reflection coefficients
on the surfaces of metallic plates in
(\ref{eq1}) are found by solving the Maxwell equations with appropriate
boundary conditions. They are
expressed via the surface impedances as \cite{67,68,69}
\begin{eqnarray}
&&
\rTM\kk=\frac{cq_l-\xi_l\zTM\kk}{cq_l+\xi_l\zTM\kk},
\nonumber \\
&&
\rTE\kk=\frac{cq_l\zTE\kk-\xi_l}{cq_l\zTM\kk+\xi_l}.
\label{eq7}
\end{eqnarray}

The surface impedances in turn are connected with the nonlocal dielectric
permittivities \cite{68,69} (see also the detailed modern rederivation in
\cite{70})
\begin{eqnarray}
&&
\zTM\kk=\frac{\xi_l}{\pi c}\int_{-\infty}^{\infty}\!\!
\frac{dk_z}{k^2}\left(\frac{c^2k_{\bot}^2}{\xi_l^2\eps_l^{L}}+
\frac{k_z^2}{\kTl^2+k_z^2}\right),
\nonumber \\
&&
\zTE\kk=\frac{\xi_l}{\pi c}\int_{-\infty}^{\infty}
\frac{dk_z}{\kTl^2+k_z^2},
\label{eq8}
\end{eqnarray}
\noindent
where $k^2=k_{\bot}^2+k_{z}^2$ and
\begin{equation}
\eps_l^{L,T}\equiv\eps^{L,T}(\ri\xi_l,\mbox{\vk}),
\quad
\kTl^2\equiv k_{\bot}^{2}+\eps_l^{T}\frac{\xi_l^2}{c^2}.
\label{eq9}
\end{equation}

If there is no dependence of the dielectric permittivities on {\vk},
one obtains
\begin{equation}
\eps^{L}(\il,0)=\eps^{T}(\il,0)=\eps_l, \quad
\kTl=k_l
\label{eq10}
\end{equation}
\noindent
and integrals in (\ref{eq8}) are easily calculated with the result
\begin{equation}
\zTM\kk=\frac{ck_l}{\xi_l\eps_l},\quad
\zTE\kk=\frac{\xi_l}{ck_l}.
\label{eq11}
\end{equation}
\noindent
Then, the substitution of (\ref{eq11}) in (\ref{eq7}) returns us back to
the standard Fresnel reflection coefficients (\ref{eq2}).

Below we suggest the phenomenological nonlocal response functions which
take dissipation into account and simultaneously bring the Lifshitz theory
in agreement with the measurement data.

\section{Phenomenological nonlocal response functions to the on-shell and
off-shell fields}

Spatially nonlocal response functions to the electromagnetic field have long
been used in the electrodynamics of solids for theoretical description of
the optical properties of charge carriers.
The nonlocal response functions $\eps^L(\omega,\mbox{\vk})$ and
$\eps^T(\omega,\mbox{\vk})$ for a collisionless electron gas have been found
in a seminal work by Lindhard \cite{71} within the random phase approximation
and generalized with account of collisions in  \cite{72}.
In the limiting case $\omega\to 0$ the obtained function  $\eps^L$
describes the Thomas-Fermi and Debye screening which has deep physical meaning
in electrostatics.
The spatially nonlocal generalizations
of the Drude response function describing the anomalous skin
effect have also been found \cite{68} using the Boltzmann equation
 and used in the Lifshitz theory
to calculate the Casimir force \cite{70,73,74}. It was  shown,
however, that at the experimental separations these nonlocal response functions
 lead to almost the same Casimir forces as the local Drude response
and do not bring theory in agreement with the measurement data.

It should be stressed that the spatially nonlocal generalizations of the
Drude response mentioned above describe the physical effects occurring in
real electromagnetic fields on the mass shell. Keeping in mind that
 an account of these effects in the Lifshitz theory does not lead
to agreement of the theoretical predictions with the measured Casimir
forces,  below we consider an alternative, phenomenological, response functions
which predict nearly the same results as the standard local response for
electromagnetic fields on the mass shell but has quite different properties for
the  off-shell fields.

Broadly speaking, the nonlocal response functions $\eps^L$ and $\eps^T$ depend on
a 3-component vector {\vk} (see the examples in \cite{71,72}).
It should be noted, however, that all the results of this type
have been obtained for the case of homogeneous and isotropic media
because in the absence of translational invariance it is impossible
in the strict sense to define
the nonlocal response functions depending on both $\omega$ and {\vk}.
Taking into consideration that in the Casimir effect the translational
invariance in the direction perpendicular to parallel plates is violated,
an immediate application of the response functions
depending on both $\omega$ and {\vk} is not warranted \cite{75,76,77}.

To illustrate our conjecture that an agreement between the Lifshitz theory
and the measurement data could be restored by modifying the response of
a metal to the off-shell fields, we consider the particular case when
 $\eps^L$ and $\eps^T$ in (\ref{eq9}) depend
not on {\vk}, but on $\kb$. This would be in line with the fact that in
the plane of the Casimir plates the translational invariance is preserved and
in direct analogy to the exact response functions of graphene which depend
just  on $\kb$ \cite{51,52,53,54}.

In this particular case, the integrals in (\ref{eq8}) are again calculated
exactly
\begin{eqnarray}
\zTM\kk&=&\frac{c\kb}{\xi_l\eps_l^L}+\frac{\xi_l}{c(k_l^T+\kb)}
\nonumber \\
&=&
\frac{c}{\xi_l}\left(\frac{\kb}{\eps_l^L}+
\frac{k_l^T-\kb}{\eps_l^T}\right),
\nonumber \\
\zTE\kk&=&\frac{\xi_l}{ck_l^T},
\label{eq12}
\end{eqnarray}
\noindent
and the substitution of these results in (\ref{eq7}) leads to the following
reflection coefficients
\begin{eqnarray}
&&
\rTM\kk=\frac{\eps_l^{T}q_l-\kTl-\kb\left(\eps_l^{T}-\eps_l^{L}
\right)\left({\eps_l^{L}}\right)^{-1}}{\eps_l^{T}q_l+\kTl+
\kb\left(\eps_l^{T}-\eps_l^{L}\right)\left({\eps_l^{L}}\right)^{-1}},
\nonumber \\
&&
\rTE\kk=\frac{q_l-\kTl}{q_l+\kTl}.
\label{eq13}
\end{eqnarray}
\noindent
With account of (\ref{eq10})
it is apparent that in the absence of spatial dispersion
(\ref{eq13}) transforms to the standard Fresnel reflection coefficients
(\ref{eq2}) commonly used in the Lifshitz theory.

In order to test a feasibility of the approach
discussed above, we consider the following alternative response functions
which present a nonlocal modification of the Drude response (\ref{eq3}):
\begin{eqnarray}
&&
\teps_{D}^{T}(\omega,\kb)=1-\frac{\omega_p^2}{\omega(\omega+\ri\gamma)}\left(
1+\ri\frac{v^{T}\kb}{\omega}\right),
\nonumber \\
&&
\teps_{D}^{L}(\omega,\kb)=1-\frac{\omega_p^2}{\omega(\omega+\ri\gamma)}\left(
1+\ri\frac{v^{L}\kb}{\omega}\right)^{-1}\!\!\!,
\label{eq14}
\end{eqnarray}
\noindent
where $v^{T,L}$ are the constants of the order of Fermi velocity $v_F$.
For $\kb=0$, the functions (\ref{eq14}) reduce to the standard Drude function
$\eps_D(\omega)=\teps_D^L(\omega,0)=\teps_D^T(\omega,0)$.

Note that the term of the order of $v_F\kb/\omega$, added to unity in
(\ref{eq14}), is the simplest dimensionless quantity which remains
negligibly small for the fields on the mass shell.
Really, for the on-shell electromagnetic field we have
\begin{equation}
\frac{v^{T,L}\kb}{\omega}\sim\frac{v_F}{c}\,\frac{c\kb}{\omega}\leq
\frac{v_F}{c}\ll 1,
\label{eq15}
\end{equation}
\noindent
i.e., any variations, as compared to the standard local Drude
response, should be only
moderate (see Sect.~5). In Sect.~6 we also prove  that the response functions
 $\teps_D^{T,L}$ are causal and satisfy the Kramers-Kronig relations (for
$\teps_D^{T}$ these relations take the form valid for functions having  the
first- and second-order poles at $\omega=0$ \cite{10}).

At the pure imaginary Matsubara frequencies the proposed nonlocal response
functions (\ref{eq14}) take the form
\begin{eqnarray}
&&
\teps_{D}^{T}\kk\equiv\teps_{D,l}^{T}=1+\frac{\omega_p^2}{\xi_l(\xi_l+\gamma)}\left(
1+\frac{v^T\kb}{\xi_l}\right),
\nonumber \\
&&
\teps_{D}^{L}\kk\equiv\teps_{D,l}^{L}=1+\frac{\omega_p^2}{\xi_l(\xi_l+\gamma)}\left(
1+\frac{v^L\kb}{\xi_l}\right)^{-1}\!\!\!\!\!\!\!.
\label{eq16}
\end{eqnarray}

Using (\ref{eq16}) and (\ref{eq13}), for the values of the reflection coefficients
at zero Matsubara frequency (this is the static limit which is off the mass shell)
one obtains
\begin{eqnarray}
&&
\rTM(0,\kb)=\frac{\omega_p^2}{\omega_p^2+2\gamma v^L\kb},
\nonumber \\
&&
\rTE(0,\kb)=\frac{\kb-\sqrt{k_{\bot}^2+\omega_p^2v^T\kb\gamma^{-1}c^{-2}}}{\kb+
\sqrt{k_{\bot}^2+\omega_p^2v^T\kb\gamma^{-1}c^{-2}}}.
\label{eq17}
\end{eqnarray}

It is seen that for $v^L=v^T=0$ the coefficients (\ref{eq17}) coincide with
(\ref{eq5}) obtained for the local Drude response. However, for different
from zero $v^L$ and $v^T$ the equations in (\ref{eq17}) are
in some sense intermediate between (\ref{eq5}) and (\ref{eq6})
related to the local Drude and plasma responses,
respectively. In so doing the contributions of $\rTM(0,\kb)$ to the
Lifshitz formula (\ref{eq1}) defined in (\ref{eq17}),
on the one hand, and in (\ref{eq5}) and (\ref{eq6}), on the other hand,
are nearly the same, whereas the contributions
of $\rTM(0,\kb)$ defined in (\ref{eq17}) and in (\ref{eq5}) are quite different.
It might be thought that theoretical predictions of the Lifshitz theory
obtained using the suggested nonlocal response functions agree with the measurement
data for the Casimir force. In the next section we provide a verification of
this assumption.

\section{Comparison between the Lifshitz theory using the alternative response
functions and experiments on measuring the Casimir force}

Before comparing experiment with theory, we compare the theoretical Casimir pressures
obtained by using the standard Drude response (\ref{eq3}), $P_D$, and its nonlocal
alternative, $\tilde{P}_D$. This comparison is made within the separation region
from 1 to $7~\mu$m between Au plates where the interband transitions do not contribute
so that the obtained results are realistic from the experimental point of view.
At first, we have calculated the values of the response functions (\ref{eq16})
at the pure imaginary Matsubara frequencies at $T=300~$K,
where for Au $\hbar\omega_p=9.0~$eV and $\hbar\gamma=35~$meV \cite{61}.
For the best agreement with the
measurement data (see below), the value $v^T=7v_F$, where for Au
$v_F=1.38\times 10^6~$m/s \cite{78}, has been chosen. The computations made by
(\ref{eq1}) and (\ref{eq13}) show that the change in the value of $v^L$
in the range from $v^L=0$
(the standard Drude response) to $v^L=10v_F$ makes only a negligibly small impact
on the values of the Casimir pressure.

\begin{figure}[b]
\vspace*{-6.5cm}
\centerline{\hspace*{3cm}
\includegraphics{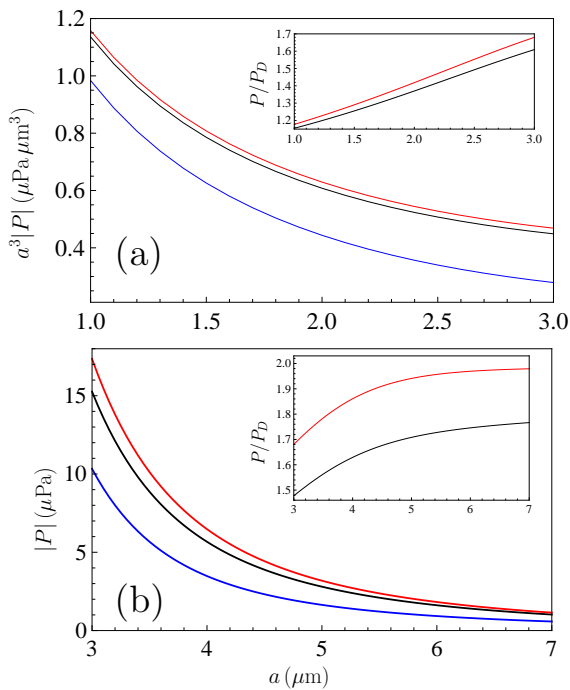}}
\vspace*{-14cm}
\caption{\label{fig1} The magnitudes of the Casimir pressure (a)
multiplied by $a^3$
and (b) on their own, computed using the standard Drude,
$P_D$, the alternative nonlocal, $\tilde{P}_D$, and the plasma,
$P_p$, responses are shown as functions of separation by the bottom, middle,
and top lines, respectively. In the insets, the ratios
$\tilde{P}_D/P_D$ and $P_p/P_D$ are shown by the bottom and top lines,
respectively.}
\end{figure}
The computational results are shown in Fig.~1 by the middle lines (a) for
$a^3|\tilde{P}_D|$ in the separation region from 1 to $3~\mu$m and
(b)  for $|\tilde{P}_D|$ in the region from 3 to $7~\mu$m as the functions
of separation between the plates.
For comparison purposes, the bottom and top lines demonstrate the respective
results $P_D$ and $P_p$ obtained using the standard Drude and plasma
response functions, given in (\ref{eq3}) and (\ref{eq4}).
In these cases the reflection coefficients (\ref{eq2}) have been used in
place of (\ref{eq13}).

As is seen in Fig.~1, the Casimir pressures obtained using the
alternative nonlocal response
are sandwiched between those found using the standard Drude and plasma responses.
Physically the differences between $P_D$ and $\tilde{P}_D$ are caused by the
fact that according to (\ref{eq17}) for the alternative nonlocal response
$\rTE(0,\kb) \neq 0$, as is also the case in (\ref{eq6}) for the plasma
response [we recall that for the local Drude function $\rTE(0,\kb)=0$
in accordance with (\ref{eq5})].
In the insets to Figs.~1(a) and 1(b),
the Casimir pressures computed using the alternative nonlocal and
the plasma responses
are normalized to $P_D$ and shown as functions of separation by
the bottom and top lines, respectively. {}From the inset to Fig.~1(b), it is
seen that at large separations of a few micrometers
the alternative nonlocal response predicts a distinctly
smaller pressure magnitudes than the plasma response.
However, in this separation region direct precise measurements of the Casimir
force are not performed yet.

Now we compare theoretical predictions of the Lifshitz theory using the
proposed alternative nonlocal response functions
with the available measurement data. For this comparison, we choose the
experiments of \cite{16} and \cite{23} where the gradient of the Casimir
force $F_{\rm expt}^{\prime}$ between a Au-coated sphere of radius $R$ and
a Au-coated plate was measured in the separation regions from 235 to 750~nm and from
0.6 to $2~\mu$m, respectively (it has been shown \cite{16} that the measurement
data of this experiment are in complete agreement with the experimental results
of earlier experiments \cite{24} obtained within the same separation region).

The Casimir pressure $P(a)$ between two Au plates given in
 (\ref{eq1}), (\ref{eq13}) and (\ref{eq16}) in the case of
alternative nonlocal response
and by (\ref{eq1})--(\ref{eq3}) for the standard one
was calculated with taken into
 account interband transitions, which occur at $\hbar\omega>2~$eV.
 The contribution of these transitions to the response function influences the
 Ca\-si\-mir pressure at separations below $1~\mu$m and their impact increases with
 decreasing separation.
 An inclusion of the interband transitions reduces to a replacement of
 the unities just after the equality sign
on the right-hand sides of (\ref{eq3}) and (\ref{eq16})
  with the respective function of the Matsubara frequencies computed
 by the standard procedure using the optical data for the complex index of
 refraction  of Au \cite{10}.
 We have also performed computations with different values of $v^T$ and made sure
that the value
 $v^T=7v_F$ leads to the best agreement between the experimental results and
 theoretical predictions (as was noted above,
the value of $v^L$ makes only a minor impact on the obtained results).

 Then, the gradient of the Casimir force between a sphere and a plate was
 calculated as \cite{23}
 \begin{eqnarray}
 &&
 F_{\rm theor}^{\prime}(a)=-2\pi R\left[1+\beta(a,R)\frac{a}{R}\right]
 \nonumber \\
 &&~~~~~~~~
 \times\left(1+10\frac{\delta_s^2+\delta_p^2}{a^2}\right)\!P(a).
 \label{eq18}
 \end{eqnarray}
 \noindent
 This equation takes into account the rms roughness $\delta_s$ and $\delta_p$ on the
 surfaces of a sphere and a plate, respectively.
The function $\beta$ takes into account the  deviations from the
 proximity force approximation used in order to adapt the Lifshitz formula (\ref{eq1})
 derived for two plates to the sphere-plate geometry
 (see \cite{16,34} for the values of all these quantities).

\begin{figure}[t]
\vspace*{-3.5cm}
\centerline{\hspace*{1.5cm}
\includegraphics{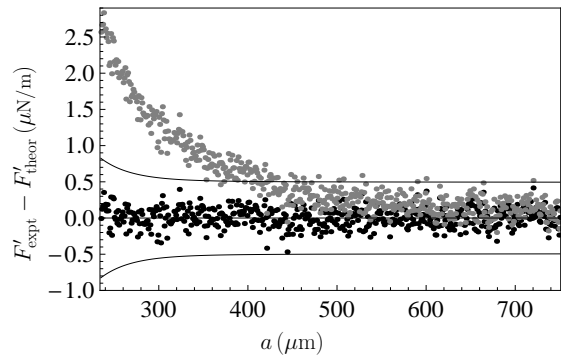}}
\vspace*{-21.3cm}
\caption{\label{fig2} The differences between experimental \cite{16}
and theoretical gradients
of the Casimir force computed using the alternative nonlocal (black dots) and
the standard Drude (grey dots) responses are shown as
functions of separation. The two solid lines indicate borders of
the 67\% confidence band. }
\end{figure}
In Fig.~\ref{fig2} the differences between $F_{\rm expt}^{\prime}$ measured in
 \cite{16} and $F_{\rm theor}^{\prime}$ computed by (\ref{eq18})
using the alternative nonlocal and the standard Drude responses are shown
by the sets of black
and grey dots, respectively. It is seen that the black dots are well inside
the 67\% confidence band shown by the two solid lines (the same is true when
the plasma response function is used in computations \cite{16}, which, however,
excludes the dissipation of free electrons from consideration). This means that the
alternative nonlocal  response is in good agreement with the measurement data.
The standard Drude response is experimentally excluded within the separation range
from 235 to 420~nm (see Fig.~\ref{fig2}).

\begin{figure}[t]
\vspace*{-6.7cm}
\centerline{\hspace*{3cm}
\includegraphics{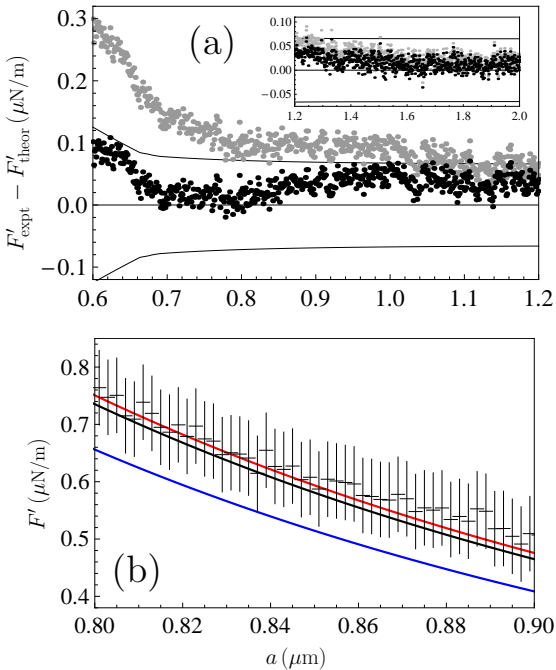}}
\vspace*{-14cm}
\caption{\label{fig3} (a) The differences
between experimental \cite{23} and theoretical
gradients of the Casimir force computed using the alternative nonlocal
(black dots) and the standard Drude (grey dots) responses are shown
as functions of separation. The two solid lines indicate borders
of the 67\% confidence band (the region of larger separations is
shown in the inset). (b) The experimental \cite{23} gradients
shown as crosses are compared with the theoretical ones computed
using the standard Drude, the alternative nonlocal, and the
plasma response functions are shown by the bottom, middle, and top lines,
respectively.}
\end{figure}

In Fig.~\ref{fig3}(a) the differences $F_{\rm expt}^{\prime}-F_{\rm theor}^{\prime}$
are shown by using the experimental data of  \cite{23} obtained at larger separations.
Once again, the Lifshitz theory using the alternative nonlocal  response (black dots)
is in agreement with the data over the entire separation region from 0.6 to $2~\mu$m
(the region of $a$ from 1.2 to $2~\mu$m is shown on the inset).
As is seen in Fig.~\ref{fig3}(a), the standard Drude response
(grey dots) is excluded in the region from 0.6 to $1.1~\mu$m.
For comparison purposes, in Fig.~\ref{fig3}(b) we also plot the experimental data as
crosses and the theoretical predictions of the Lifshitz theory using the
standard Drude, the alternative nonlocal, and the plasma response functions as the
bottom, middle and top lines, respectively. As is seen in Fig.~\ref{fig3}(b), the
theoretical predictions  are in agreement with the measurement data when
using the alternative nonlocal  or the plasma responses,
but are excluded by the same data if the standard Drude response is used.
Once again, the alternative nonlocal response can be considered as preferential
because it takes into account the dissipation of free electrons which is disregarded
by the plasma response function.

\section{The alternative nonlocal  response and reflection of electromagnetic waves on
the mass shell}

According to the above results, the alternative nonlocal
 response functions introduced
in Sect.~3 bring the Lifshitz theory in agreement with measurements of the Casimir
force which was unattainable when using the standard (spatially local) Drude
response. This has been made possible due to a peculiarity of nonlocal responses to the
fluctuating fields off the mass shell. Below we check that the
alternative nonlocal  functions describe
correctly the response of metal to usual electromagnetic waves on the mass shell.

For this purpose, we consider the electromagnetic wave on the mass shell incident under
an angle $\theta$ on an Au plate described by the nonlocal response functions (\ref{eq14}).
Using the relationship $\sin\theta=\kb c/\omega$, one can present the reflection
coefficients (\ref{eq13}) calculated along the real frequency axis in the following form:
\begin{eqnarray}
&&
r_{\rm TM}(\omega,\theta)=
\frac{\teps_{D}^{T}\cos\theta-\sqrt{\teps_{D}^{T}-\sin^2\theta}+
\ri\frac{\sin\theta(\teps_{D}^{T}-\teps_{D}^{L})}{\teps_{D}^{L}}}{\teps_{D}^{T}\cos\theta
+\sqrt{\teps_{D}^{T}-\sin^2\theta}-
\ri\frac{\sin\theta(\teps_{D}^{T}-\teps_{D}^{L})}{\teps_{D}^{L}}},
\nonumber \\
&&
r_{\rm TE}(\omega,\theta)=
\frac{\cos\theta-\sqrt{\teps_{D}^{T}-\sin^2\theta}}{\cos\theta
+\sqrt{\teps_{D}^{T}-\sin^2\theta}},
\label{eq19}
\end{eqnarray}
\noindent
where now
$\teps_{D}^{T,L}=\teps_{D}^{T,L}(\omega,\theta)$.
According to (\ref{eq14}), one obtains
\begin{eqnarray}
&&
\teps_{D}^{T}(\omega,\theta)=1-\frac{\omega_p^2}{\omega(\omega^2+\gamma^2)}
\left[\omega+\gamma\frac{v^T}{c}\sin\theta\right.
\nonumber \\
&&~~~~~~~~~
+\left.
\ri\left(\gamma-\omega\frac{v^T}{c}\sin\theta\right)\right],
\nonumber \\
&&
\teps_{D}^{L}(\omega,\theta)=1-\frac{\omega_p^2}{\omega(\omega^2+\gamma^2)}
\left[\omega-\gamma\frac{v^L}{c}\sin\theta
\right.
\nonumber \\
&&~~~~~~~~
+\left.
\ri\left(\gamma+\omega\frac{v^L}{c}\sin\theta\right)\right]
\left(1+\frac{{v^L}^2}{c^2}\sin^2\theta\right)^{-1}\!\!\!\!\!\!\!.
\label{eq20}
\end{eqnarray}

We have computed the reflectances defined by the alternative nonlocal functions
\begin{eqnarray}
&&
{\cal R}_{\rm TM}(\omega,\theta)=|\rTM(\omega,\theta)|^2,
\nonumber \\
&&
{\cal R}_{\rm TE}(\omega,\theta)=|\rTE(\omega,\theta)|^2,
\label{eq21}
\end{eqnarray}
\noindent
where $\rTM$ and $\rTE$ are defined in (\ref{eq19}) and (\ref{eq20}) with
$v^L=v^T=7v_F$, over the frequency region $\hbar\omega$ from 0.1 to 1~eV.
In this frequency region, the standard Drude response function is usually used for the
interpretation of measured optical data for the complex index of refraction.

We have also computed the standard Drude ref\-lec\-tan\-ces
\begin{eqnarray}
&&
{\cal R}_{{\rm TM},D}(\omega,\theta)=|r_{{\rm TM},D}(\omega,\theta)|^2,
\nonumber \\
&&
{\cal R}_{{\rm TE},D}(\omega,\theta)=|r_{{\rm TE},D}(\omega,\theta)|^2,
\label{eq21}
\end{eqnarray}
\noindent
which are also given by (\ref{eq19}) and (\ref{eq20}), but with
$v^L=v^T=0$.

The relative deviations between the reflectances obtained by using the
alternative nonlocal and standard Drude response functions are
\begin{equation}
\delta{\cal R}_{\rm TM}(\omega,\theta)=\frac{{\cal R}_{\rm TM}(\omega,\theta)-
{\cal R}_{{\rm TM},D}(\omega,\theta)}{{\cal R}_{{\rm TM},D}(\omega,\theta)}
\label{eq23}
\end{equation}
\noindent
for the TM electromagnetic waves and
\begin{equation}
\delta{\cal R}_{\rm TE}(\omega,\theta)=\frac{{\cal R}_{\rm TE}(\omega,\theta)-
{\cal R}_{{\rm TE},D}(\omega,\theta)}{{\cal R}_{{\rm TE},D}(\omega,\theta)}
\label{eq24}
\end{equation}
\noindent
for the TE ones.

\begin{figure}[t]
\vspace*{-5.9cm}
\centerline{\hspace*{3cm}
\includegraphics{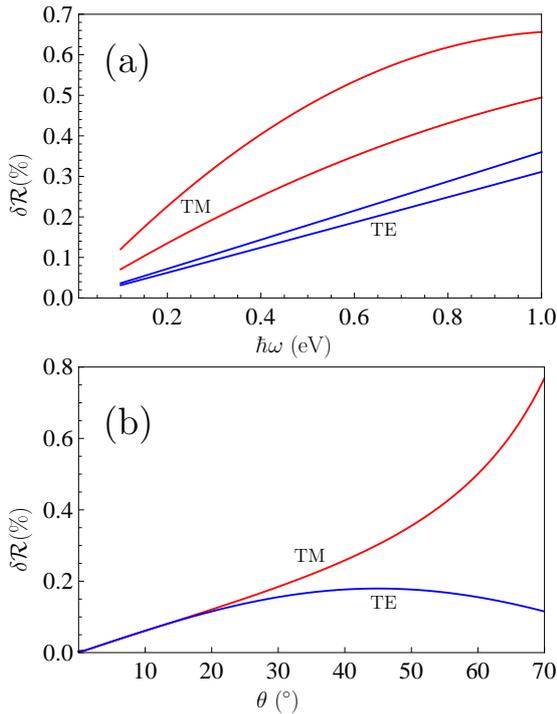}}
\vspace*{-14.2cm}
\caption{\label{fig4} Relative deviations between the reflectances
of electromagnetic waves on the mass shell
incident on an Au plate under the angle $\theta$, which are
computed using the alternative nonlocal  and standard Drude responses,
 are shown as (a) the functions of frequency
by the top and bottom pairs of lines  for the TM and TE polarizations,
respectively (the upper and lower lines in the top pair are for
$\theta=\pi/3$ and $\pi/4$, respectively, and vice versa for the bottom
pair) and (b) the functions of the incidence angle at $\hbar\omega=0.5~$eV
by the top and bottom lines  for the TM and TE polarizations,
respectively.}
\end{figure}

The computational results for $\delta{\cal R}_{\rm TM}$ and $\delta{\cal R}_{\rm TE}$
are presented in Fig.~\ref{fig4}. The top and bottom pairs of lines in  Fig.~\ref{fig4}(a)
show $\delta{\cal R}_{\rm TM}$ and $\delta{\cal R}_{\rm TE}$, respectively, as the
functions of $\hbar\omega$. The upper and lower lines in the top pair are for
$\theta=\pi/3$ and $\pi/4$ angles of incidence, respectively.
In the bottom pair of lines, the upper line is for $\theta=\pi/4$ and the lower line
for $\theta=\pi/3$. The dependences of $\delta{\cal R}_{\rm TM}$ (the top line)
and $\delta{\cal R}_{\rm TE}$ (the bottom line) on the incidence angle are illustrated
in Fig.~\ref{fig4}(b) at $\hbar\omega=0.5~$eV.

As is seen in Fig.~\ref{fig4}, the relative deviations between the reflectances
calculated using the alternative nonlocal  and the standard Drude response
functions do not exceed a fraction of a percent. This is below experimental
 errors in measuring the optical data for the complex index of refraction and
in determination of the parameters $\omega_p$ and $\gamma$ in the Drude response.
Thus, the suggested alternative nonlocal response  leads to the same
experimental consequences for the electromagnetic waves on the mass shell
as the standard Drude one.
In regard to fluctuating fields off the mass shell, the electromagnetic response to them,
as discussed in Sect.~1, cannot be immediately measured. Some further
circumstantial evidence about it can be obtained only concerning $\eps^L$
\cite{67} which, however, does not affect a comparison between the Lifshitz
theory and the measurement data (see Sect.~4).

\section{The principle of causality and the Kramers-Kronig relations for the
alternative response functions}
\def\Xint#1{\mathchoice
   {\XXint\displaystyle\textstyle{#1}}%
   {\XXint\textstyle\scriptstyle{#1}}%
   {\XXint\scriptstyle\scriptscriptstyle{#1}}%
   {\XXint\scriptscriptstyle\scriptscriptstyle{#1}}%
   \!\int}
\def\XXint#1#2#3{{\setbox0=\hbox{$#1{#2#3}{\int}$}
     \vcenter{\hbox{$#2#3$}}\kern-.5\wd0}}
\def\ddashint{\Xint=}
\def\dashint{\Xint-}

As shown in Sects.~4 and 5, the proposed alternative nonlocal response
functions not only bring the Lifshitz theory in agreement with measurements of the
Casimir force (this has been made possible due to a modified contribution from the
off-shell fields), but is also in good agreement with well established
physics determined by the on-shell electromagnetic waves.
It is pertinent now to make sure that the proposed nonlocal response functions
satisfy the fundamental principle of causality formulated mathematically as the
Kramers-Kronig relations. According to this principle, the response function in
the $(\mbox{\boldmath$x$},t)$ representation must be determined by the field values only
at the previous moments $t^{\prime}<t$ and, as a result, its Fourier image must be
an analytic function in the upper half plane of complex $\omega$ \cite{66}.

We begin with the transverse response function $\teps_D^T$ defined in the first
line of (\ref{eq14}). The real and imaginary parts of $\teps_D^T$ defined along
the real frequency axis are given by
\begin{equation}
{\rm Re\,}\teps_D^T(\omega,\kb)=1-
\frac{\omega_p^2(\omega^2+\gamma v^T\kb)}{\omega^2(\omega^2+\gamma^2)}
\label{eq25}
\end{equation}
\noindent
and
\begin{equation}
{\rm Im\,}\teps_D^T(\omega,\kb)=
\frac{\omega_p^2(\gamma -v^T\kb)}{\omega(\omega^2+\gamma^2)}.
\label{eq26}
\end{equation}
\noindent
{}From (\ref{eq26}) it is seen that ${\rm Im\,}\teps_D^T$ has the pole of the first
order, as it holds for the standard Drude function (\ref{eq3}), whereas
${\rm Re\,}\teps_D^T$ in (\ref{eq25}) possesses the pole of the second order at
zero frequency similar to the plasma response function (\ref{eq4}).

It has been known \cite{66,79} that the form of Kra\-mers-Kronig relations depends on
whether the dielectric permittivity is regular at $\omega=0$ or it has the poles of the
first or second order. The dielectric permittivity $\teps_D^T$ is the sum of two
functions (\ref{eq25}) and (\ref{eq26}) having the first- and second-order poles at
$\omega=0$. Because of this, the standard Kramers-Kronig relations derived for regular at
$\omega=0$ functions undergo two respective modifications.

At first, we consider the
Kramers-Kronig relation expressing the real part of the dielectric permittivity
$\teps_D^T$ via its imaginary part.
The form of this relation is not influenced by the presence of the
first-order pole at $\omega=0$ \cite{66}.
However, as it was shown previously \cite{79} in a more simple case of the plasma
response function (\ref{eq4}), due to the presence of the second-order pole in
${\rm Re\,}\teps_D^T$, one should consider the quantity
\begin{equation}
F(\omega,\kb)\equiv 1+\frac{1}{\pi}\dashint_{-\infty}^{\infty}\!\!dx
\frac{{\rm Im\,}\teps_D^T(x,\kb)}{x-\omega}-\frac{\omega_p^2}{\omega^2}
\frac{v^T\kb}{\gamma}.
\label{eq27}
\end{equation}
\noindent
The term subtracted on the right-hand side of this equation represents the
behavior of ${\rm Re\,}\teps_D^T$ in the vicinity of $\omega=0$
(the integrals here and below are understood as the principal values).

Substituting (\ref{eq26}) in (\ref{eq27}), one obtains
\begin{eqnarray}
&&
F(\omega,\kb)= 1+\frac{\omega_p^2(\gamma-v^T\kb)}{\pi}
\dashint_{-\infty}^{\infty}\!
\frac{dx}{x(x^2+\gamma^2)(x-\omega)}
\nonumber \\
&&~~~~~~~~~~~~
-\frac{\omega_p^2}{\omega^2}
\frac{v^T\kb}{\gamma}.
\label{eq28}
\end{eqnarray}
\noindent

Then, calculating the integrals on the right-hand side of (\ref{eq28}), we arrive at
\begin{eqnarray}
&&
F(\omega,\kb)= 1+\frac{\omega_p^2(\gamma-v^T\kb)}{\pi}\left[
-\frac{1}{\omega\gamma^2}\dashint_{-\infty}^{\infty}\!\frac{dx}{x}
\right.
\nonumber \\
&&
+\frac{1}{\omega(\omega^2+\gamma^2)}\dashint_{-\infty}^{\infty}\!\frac{dx}{x-\omega}
+\frac{\omega}{\gamma^2(\omega^2+\gamma^2)}
\dashint_{-\infty}^{\infty}\!\frac{x\,dx}{x^2+\gamma^2}
\nonumber \\
&&\left.
-\frac{1}{\omega^2+\gamma^2}
\dashint_{-\infty}^{\infty}\!\frac{dx}{x^2+\gamma^2}\right]
-\frac{\omega_p^2}{\omega^2}\frac{v^T\kb}{\gamma}.
\label{eq29}
\end{eqnarray}
\noindent
Taking into account that the first three integrals here are equal to zero and
calculating the fourth one, we find
\begin{eqnarray}
&&
F(\omega,\kb)= 1-\frac{\omega_p^2(\gamma-v^T\kb)}{(\omega^2+\gamma^2)\gamma}
-\frac{\omega_p^2}{\omega^2}\frac{v^T\kb}{\gamma}
\nonumber \\
&&~~~
=1-\frac{\omega_p^2(\omega^2+\gamma v^T\kb)}{\omega^2(\omega^2+\gamma^2)}
={\rm Re\,}\teps_D^T(\omega,\kb).
\label{eq30}
\end{eqnarray}

Thus, the first Kramers-Kronig relation for the alternative nonlocal response function
$\teps_D^T$ takes the form
\begin{eqnarray}
&&
{\rm Re\,}\teps_D^T(\omega,\kb)=
1+\frac{1}{\pi}\dashint_{-\infty}^{\infty}\!\!dx
\frac{{\rm Im\,}\teps_D^T(x,\kb)}{x-\omega}
\nonumber\\
&&~~~~~~~~~~~~~~~~~~
-\frac{\omega_p^2}{\omega^2}\frac{v^T\kb}{\gamma}.
\label{eq31}
\end{eqnarray}

Next, we express the imaginary part of $\teps_D^T$ via its real part.
For this purpose we consider the quantity
\begin{eqnarray}
&&
G(\omega,\kb)\equiv-\frac{1}{\pi}\dashint_{-\infty}^{\infty}\!
\frac{dx}{x-\omega}\left[{\rm Re\,}\teps_D^T(x,\kb)+
\frac{\omega_p^2}{x^2}\frac{v^T\kb}{\gamma}\right]
\nonumber \\
&&~~~~~~~~~~~~~~~~~
+\frac{\omega_p^2}{\omega\gamma^2}\left(\gamma-v^T\kb\right).
\label{eq32}
\end{eqnarray}

The second term in square brackets of (\ref{eq32})
arises because in the vicinity of $\omega=0$
${\rm Re\,}\teps_D^T$ has the second-order pole. For dielectric functions
possessing the second-order pole at $\omega=0$, the presence of such a term in
the second Kramers-Kronig relation was proven in \cite{79} by the example of
the plasma response function (\ref{eq4}).

The last term on the right-hand side of (\ref{eq32}) represents the behavior
of ${\rm Im\,}\teps_D^T$ in the vicinity of the first-order pole. The term of
this kind is present in the se\-cond Kramers-Kronig relation for metals \cite{66}
and can be interpreted in terms of the transverse conductivity defined as
\begin{equation}
\teps_D^T(\omega,\kb)=1+\ri\frac{4\pi\tilde{\sigma}_D^T(\omega,\kb)}{\omega}.
\label{eq33}
\end{equation}
\noindent
Using (\ref{eq26}), it is easily seen that
\begin{equation}
{\rm Re\,}\tilde{\sigma}_D^T(\omega,\kb)=
\frac{\omega_p^2(\gamma-v^T\kb)}{4\pi(\omega^2+\gamma^2)}
\label{eq34}
\end{equation}
\noindent
and, thus, in the static limit,
\begin{equation}
{\rm Re\,}\tilde{\sigma}_{D,0}^T(\kb)=
\lim_{\omega\to 0}{\rm Re\,}\tilde{\sigma}_D^T(\omega,\kb)=
\frac{\omega_p^2(\gamma-v^T\kb)}{4\pi\gamma^2}.
\label{eq35}
\end{equation}

In the local limit $\kb\to 0$, (\ref{eq35}) transforms into the static
conductivity of the standard Drude response function
\begin{equation}
\lim_{\kb\to 0}{\rm Re\,}\tilde{\sigma}_{D,0}^T(\kb)\equiv\sigma_{D,0}
=\frac{\omega_p^2}{4\pi\gamma}.
\label{eq36}
\end{equation}

Substituting (\ref{eq25}) in (\ref{eq32}), one obtains
\begin{eqnarray}
&&
G(\omega,\kb)=-\frac{1}{\pi}\dashint_{-\infty}^{\infty}\!
\frac{dx}{x-\omega}\left[1-
\frac{\omega_p^2(x^2+\gamma v^T\kb)}{x^2(x^2+\gamma^2)}
\right.
\nonumber \\
&&~~~~~~~~~~
+\left.
\frac{\omega_p^2}{x^2}\frac{v^T\kb}{\gamma}\right]
+\frac{\omega_p^2}{\omega\gamma^2}\left(\gamma-v^T\kb\right).
\label{eq37}
\end{eqnarray}
\noindent
Calculating the integrals on the right-hand side of this equation and again
omitting that ones equal to zero, we arrive at
\begin{eqnarray}
&&
G(\omega,\kb)=-\frac{\omega_p^2\omega(\gamma -v^T\kb)}{\gamma^2(\omega^2+\gamma^2)}
+\frac{\omega_p^2}{\omega\gamma^2}\left(\gamma-v^T\kb\right)
\nonumber \\
&&~~~~~~~~~
=\frac{\omega_p^2(\gamma -v^T\kb)}{\omega(\omega^2+\gamma^2)}
={\rm Im\,}\teps_D^T(\omega,\kb)
\label{eq38}
\end{eqnarray}
\noindent
in accordance with (\ref{eq26}).

Thus, the second Kramers-Kronig relation for the permittivity $\teps_D^T$ has
the form
\begin{eqnarray}
&&
{\rm Im\,}\teps_D^T(\omega,\kb)=
-\frac{1}{\pi}\dashint_{-\infty}^{\infty}\!
\frac{dx}{x-\omega}
\nonumber \\
&&~~~~~~~~~~~~~~~~~
\times\left[{\rm Re\,}\teps_D^T(x,\kb)+
\frac{\omega_p^2}{x^2}\frac{v^T\kb}{\gamma}\right]
\nonumber \\
&&~~~~~~~~~~~~~~~
+\frac{4\pi{\rm Re\,}\tilde{\sigma}_{D,0}^T(\kb)}{\omega},
\label{eq39}
\end{eqnarray}
\noindent
where ${\rm Re\,}\tilde{\sigma}_{D,0}^T$ is defined in (\ref{eq35}).

In order to derive the Kramers-Kronig relation for the dielectric permittivity
$\teps_D^T$ along the imaginary frequency axis, we consider the quantity
\begin{equation}
H(\xi,\kb)\equiv 1+\frac{2}{\pi}\int_{0}^{\infty}\!\!\!\!\!\!dx
\frac{x{\rm Im\,}\teps_D^T(x,\kb)}{x^2+\xi^2}+\frac{\omega_p^2}{\xi^2}
\frac{v^T\kb}{\gamma}.
\label{eq40}
\end{equation}
\noindent
The last term on the right-hand side of this equation should be added because
${\rm Re\,}\teps_D^T$ possesses the second-order pole at zero frequency \cite{79}.
Substituting (\ref{eq26}) in (\ref{eq40}), one obtains
\begin{eqnarray}
&&
H(\xi,\kb)= 1+\frac{2\omega_p^2(\gamma-v^T\kb)}{\pi}\!\!\int_{0}^{\infty}\!\!\!\!
\frac{dx}{(x^2+\xi^2)(x^2+\gamma^2)}
\nonumber \\
&&~~~~~~~~~~~~~~~~
+\frac{\omega_p^2}{\xi^2}
\frac{v^T\kb}{\gamma}.
\label{eq41}
\end{eqnarray}

Now we calculate the integrals on the right-hand side of this equation and obtain
\begin{eqnarray}
&&
H(\xi,\kb)= 1+\frac{\omega_p^2(\gamma-v^T\kb)}{(\xi+\gamma)\xi\gamma}
+\frac{\omega_p^2}{\xi^2}\frac{v^T\kb}{\gamma}
\nonumber \\
&&~~~~~~~
=1+\frac{\omega_p^2}{\xi^2}\,\frac{\xi+v^T\kb}{\xi+\gamma}
=\teps_D^T(\ri\xi,\kb)
\label{eq42}
\end{eqnarray}
in accordance to the first line in (\ref{eq16}). From (\ref{eq40}) and (\ref{eq42})
we finally find
\begin{equation}
\teps_D^T(\ri\xi,\kb)= 1+\frac{2}{\pi}\!\int_{0}^{\infty}\!\!\!\!\!dx
\frac{x{\rm Im\,}\teps_D^T(x,\kb)}{x^2+\xi^2}+\frac{\omega_p^2}{\xi^2}
\frac{v^T\kb}{\gamma}.
\label{eq43}
\end{equation}
\noindent
This equation has the same form as was derived in \cite{79} for the generalized
plasma-like response function taking into account the interband transitions in the
framework of the oscillator model. As was noted in \cite{66}, the presence of the
first-order pole makes no impact on the Kramers-Kronig relation expressing the
dielectric permittivity along the imaginary frequency axis.

We continue with the longitudinal alternative nonlocal response function defined
in the second line of (\ref{eq14}). This function can be equivalently written in
the form
\begin{equation}
\teps_D^L(\omega,\kb)=1-\frac{\omega_p^2}{(\omega+\ri\gamma)(\omega+\ri v^L\kb)},
\label{eq44}
\end{equation}
\noindent
i.e. it is an analytic function in the upper half plane of complex frequencies
including the entire real frequency axis. Because of this, the permittivity
$\teps_D^L$ satisfies the standard Kramers-Kronig relations derived for
insulators \cite{66}.

{}From (\ref{eq44}) one finds
\begin{eqnarray}
&&
{\rm Re\,}\teps_D^L(\omega,\kb)=1-
\frac{\omega_p^2(\omega^2-\gamma v^L\kb)}{(\omega^2+\gamma^2)
\left(\omega^2+ {v^L}^2k_{\bot}^2\right)},
\nonumber \\
&&
{\rm Im\,}\teps_D^L(\omega,\kb)=
\frac{\omega\omega_p^2(\gamma +v^L\kb)}{(\omega^2+\gamma^2)
\left(\omega^2+ {v^L}^2k_{\bot}^2\right)}.
\label{eq45}
\end{eqnarray}

Direct calculation using (\ref{eq45}) results in the familiar relations
\begin{eqnarray}
&&
{\rm Re\,}\teps_D^L(\omega,\kb)=1+\frac{1}{\pi}\dashint_{-\infty}^{\infty}\!\!\!
dx\frac{{\rm Im\,}\teps_D^L(x,\kb)}{x-\omega},
\nonumber \\
&&
{\rm Im\,}\teps_D^L(\omega,\kb)=-\frac{1}{\pi}\dashint_{-\infty}^{\infty}\!\!\!
dx\frac{{\rm Re\,}\teps_D^L(x,\kb)}{x-\omega}.
\label{eq46}
\end{eqnarray}

The following equality is also valid:
\begin{equation}
\teps_D^L(\ri\xi,\kb)=1+\frac{2}{\pi}\int_{0}^{\infty}\!\!\!
dx\frac{x{\rm Im\,}\teps_D^L(x,\kb)}{x^2+\xi^2}.
\label{eq47}
\end{equation}

Thus, being the analytic functions
in the upper half plane of complex frequencies, the nonlocal
alternatives $\teps_D^T$ and $\teps_D^L$ to the Drude response
function $\eps_D$ considered in this paper
are causal and satisfy the Kramers-Kronig relations.
The specific form of these relations found above depends on a behavior of the
response function in the vicinity of zero frequency.

\section{Conclusions and discussion}

In the foregoing, we have proposed the phenomenological spatially
nonlocal response functions to the
electromagnetic field which are alternative to the standard Drude
function. Unlike nonlocal responses already
considered in the literature (see Sect. 1), the suggested
ones lead to nearly the same results, as the standard
Drude response, in the range of electromagnetic fields and
fluctuations on the mass shell, but cause significant differences
for the off-shell fields.

We have demonstrated that theoretical predictions of the Lifshitz
theory using the optical data of Au extrapolated down to zero
frequency by means of the proposed nonlocal response functions are
in a very good agreement with the experimental data on measuring
the Casimir force. This can be considered as a step forward in
resolution of the Casimir puzzle discussed in Sect.~1. The key
advantage of the alternative nonlocal response is that it takes into
account the dissipation of conduction electrons, whereas previously
it was necessary to simply discard it in order to bring the
Lifshitz theory in agreement with the measurement data. It was also
shown that the suggested response functions lead to nearly the same
reflectances of the electromagnetic waves on the mass shell as the
standard Drude response. This confirms that the dissipation of free
electrons is properly accounted for by the alternative electromagnetic
response proposed here.

As mentioned in Sect.~1, the spatially nonlocal alternatives to
the Drude function were also found for graphene whose
electromagnetic response described by the polarization tensor was
derived on the basis of first principles of thermal quantum field
theory. As discussed in Sect.~1, the Lifshitz theory using the
exact response functions of graphene is in good agreement with
measurements of the Casimir force in graphene systems. The same
response functions considered on the mass shell describe correctly
the reflectances of graphene \cite{53,80} and satisfy the
Kramers-Kronig relations \cite{82}.

The suggested here alternative nonlocal response functions offer a similar
situation for Au plates. It is remarkable also that, much like
graphene described by the polarization tensor, the Casimir entropy
for metallic plates described by these response functions
satisfy the Nernst heat theorem. This is eventually connected with
the fact that the value of the obtained TE reflection
coefficient at zero frequency is not equal to zero [see the second
line in (17)], rather than for the standard Drude function where it
vanishes (the detailed proof of this statement will be provided
elsewhere).

It may be argued that the alternative nonlocal response functions
suggested above are somewhat arbitrary and cannot be considered as a
rigorous solution of the problem. It is true that these
functions are introduced phenomenologically as the simplest example
of a situation where the responses to electromagnetic fluctuations
on and off the mass shell are dissimilar. Taking into consideration,
however, that after twenty years of much work on the subject made by
many researchers an agreement between theoretical predictions for the
Casimir force with account of dissipation and precise measurements
has not been reached, an employment of the phenomenological approach
can be considered as warranted.

To conclude, even though the proposed nonlocal response functions do
not provide a final resolution of the Casimir puzzle, they may point
the new way to its resolution.

\begin{acknowledgement}
This work was supported by the Peter the Great
Saint Petersburg Polytechnic University in the framework
of the Program ``5--100--2020".
V.M.M.~was partially funded by the Russian Foundation for Basic
Research, grant number 19-02-00453 A.
V.M.M.\ was also partially supported by the Russian Government
Program of Competitive Growth of Kazan Federal University.
\end{acknowledgement}

\end{document}